\documentclass[nofootinbib]{revtex4}
\usepackage[T1]{fontenc}
\usepackage{amsmath,amssymb}
\usepackage{epsfig}
\usepackage{dcolumn}
\usepackage{graphicx}
\usepackage[usenames,dvipsnames]{color}
\usepackage{slashed}
\usepackage[colorlinks,citecolor=blue]{hyperref}
\usepackage{pdfpages}
\usepackage{float}
\usepackage{adjustbox}
\usepackage[autostyle]{csquotes}
\begin{document}
\title{\boldmath Uncertainties in neutrino oscillation parameter sensitivity due to resonance processes at NO$\nu$A}
\author{Paramita Deka \footnote{paramitadeka@gauhati.ac.in}, Nilavjyoti Hazarika \footnote{nilavhazarika@gauhati.ac.in}, Kalpana Bora \footnote{kalpana@gauhati.ac.in}} 

\affiliation{Department Of Physics, Gauhati University, Assam, India}

\begin{abstract}

\textbf{Abstract}\\

The long baseline (LBL) neutrino experiments use heavy nuclear targets for neutrino scattering in which nuclear effects give rise to complications in measuring the neutrino oscillation parameters up to high precision. These nuclear effects are not yet fully understood and therefore need to be quantified as they contribute to the systematic uncertainties. Precision reconstruction of neutrino energy is one of the main components in measuring the oscillation parameters, and it is required that the neutrino energy is reconstructed with very high precision. In this work, we investigate the effects of the resonance (RES) interaction process using two models for carbon target for $\nu_{\mu}\rightarrow\nu_{\mu}$ disappearance channel of the NO$\nu$A experiment, on neutrino-nucleus scattering cross section, events, and neutrino oscillation parameter sensitivity. We also incorporate the realistic detector specifications of NO$\nu$A. To quantify the systematic uncertainties due to RES interactions, we compare the cross-sections, events, and sensitivity analysis for two models of RES processes - Rein-Sehgal and Berger-Sehgal, and comment on which model produces more precision. We observe that RES processes contribute significantly, and should be included carefully in models while extracting neutrino oscillation parameters.\\


\end{abstract}

\maketitle

\section{\label{sec:level1}Introduction}

Neutrino oscillation experiments such as T2K \cite{Abe:2017vif}, MINOS \cite{Michael:2006rx}, NO$\nu$A \cite{NOvA:2016vij}, DUNE (formerly known as LBNE) \cite{DUNE:2020txw} give/are expected to give novel information on neutrino oscillation parameters-mixing angles, mass hierarchies and CP violating phase. In these experiments, the neutrino flux at the far detector (FD) is compared with the flux at the near detector (ND), and from this one can determine the neutrino oscillation parameters. In the real-time scenario, we compare the event rate i.e., flux times cross-section at a given neutrino energy $E_{\nu}$ at the FD with that at the ND. For comparison of the event rates, a precise value of incoming neutrino energy is needed. The main complication arises from the fact that we do not know the neutrino energy a priori as the neutrinos are produced as the secondary decay products (mostly pions and kaons) from primary interactions of protons with nuclei. Therefore, the neutrino energy should be reconstructed on an event-by-event basis from the final state particles. For neutrino energy reconstruction either of the two methods are used - (i) Kinematic method \cite{Coloma:2013tba} and (ii) Calorimetric method \cite{Ankowski:2015jya, Ankowski:2015kya}. In the kinematic method, the energy of the incoming neutrino can be calculated from the kinematics (energy and angle) of the outgoing lepton for a charge current (CC) quasi-elastic (QE) scattering of a neutrino on a free nucleon at rest. On the other hand, the calorimetric method estimates the neutrino energy by summing all the energies of the particles in the final state. As modern experiments use heavy nuclear targets they suffer from complications of nuclear effects. Firstly, the neutron is not free and Fermi moving inside a nuclear potential well. Secondly, due to final state interactions (FSIs), non-QE events may be wrongly identified as QE events. These events are reconstructed with energy that is not correct. At lower energies, QE and pion production plays a significant role while Fermi motion produces a distribution of reconstructed energy around the true neutrino energy. In some cases, $\Delta$ resonance is produced in the initial neutrino-nucleus interaction vertex, e.g., $\nu p\rightarrow \mu^{-}\Delta^{++}$, and then $\Delta$ is reabsorbed in the nucleus through FSIs. The $\Delta$s can also decay to pions, and these pions are not visible in the final state and such an event is considered a QE-like event even though initially it is not true QE origin. In any QE interaction pion production is one of the major sources of background as CC $\pi^{+}$ produces a background in $\nu_{\mu}$ disappearance experiments while neutral current (NC) $\pi^{0}$ production in $\nu_{e}$ appearance experiments. Therefore it is very important to understand the pion production mechanism to minimize the systematic uncertainties in neutrino interaction cross-sections and measurement of oscillation parameters. The neutrino-induced delta resonance and pion productions result from nuclear effects. Major nuclear effects present are uncertainties from the binding energy, multi-nucleon correlation, nuclear Fermi motion effects, and FSI of produced hadrons in various interaction processes \cite{Devi:2020lem}. Uncertainties that arise due to the presence of nuclear effects are one of the leading sources of systematic errors, and hence to reduce the systematic uncertainties, the neutrino-nucleus cross-sections should be measured and known precisely.\\

For neutrino energy, $\sim$ 1-2 GeV, QE interaction is the dominant CC process for neutrino-nucleus interactions and it is the relevant neutrino energy for ongoing LBL neutrino oscillation experiments \cite{Formaggio:2013kya}. The QE interaction process is used as a signal event in many neutrino oscillation experiments operating in this energy region. Neutrino-induced single pion production (SPP) has a significant contribution at low energy in the range of 0.5-3 GeV, which is the main background for CCQE events. As neutrino energy increases, resonance pion production (RES) and Deep Inelastic Scattering (DIS) become more active and they may lead to many hadrons in the final state which again lead to the wrong determination of neutrino energy reconstruction. The NO$\nu$A neutrino flux peaks around 2 GeV and in this region a significant contribution comes from resonance productions. In RES processes, neutrinos with enough energy can excite the struck nucleon into a resonant state. These resonant states are unstable which further decay into mesons and nucleons along with a charged lepton in case of CC interaction - $\nu_{\mu}N\rightarrow\nu^{-}N^{*}$, $N^{*}\rightarrow\pi N^{\prime}$ where N, $N^{\prime}$= n, p. The dominant contribution comes from the SPP mediated by the $\Delta$ (1232) resonance. The RES process includes - the baryon resonance decay, the absorption of pion during FSI inside the nucleus, and the charge exchange of a neutral pion inside the nucleus. Among these, FSI is one of the most relevant nuclear effects which produces fake events. The lower energy neutrino experiments like MiniBooNE and MicroBooNE are mainly sensitive to QE and resonance interactions. In these experiments, the contribution of pion production is one-third of the total cross-section while for the higher neutrino energy experiments such as NO$\nu$A, MINER$\nu$A, DUNE this contribution increases to two-thirds \cite{Mosel:2012kt}. Therefore, it is crucial to study and understand this interaction channel. To get a precise knowledge of nuclear effects to control the systematic uncertainties, current studies are not enough. In these experiments argon and carbon are extensively used as a target, hence this study will help to quantify the systematics associated with RES processes in the development of models.\\

\begin{figure}[H]
\centering\includegraphics[width=13.5cm, height=8cm]{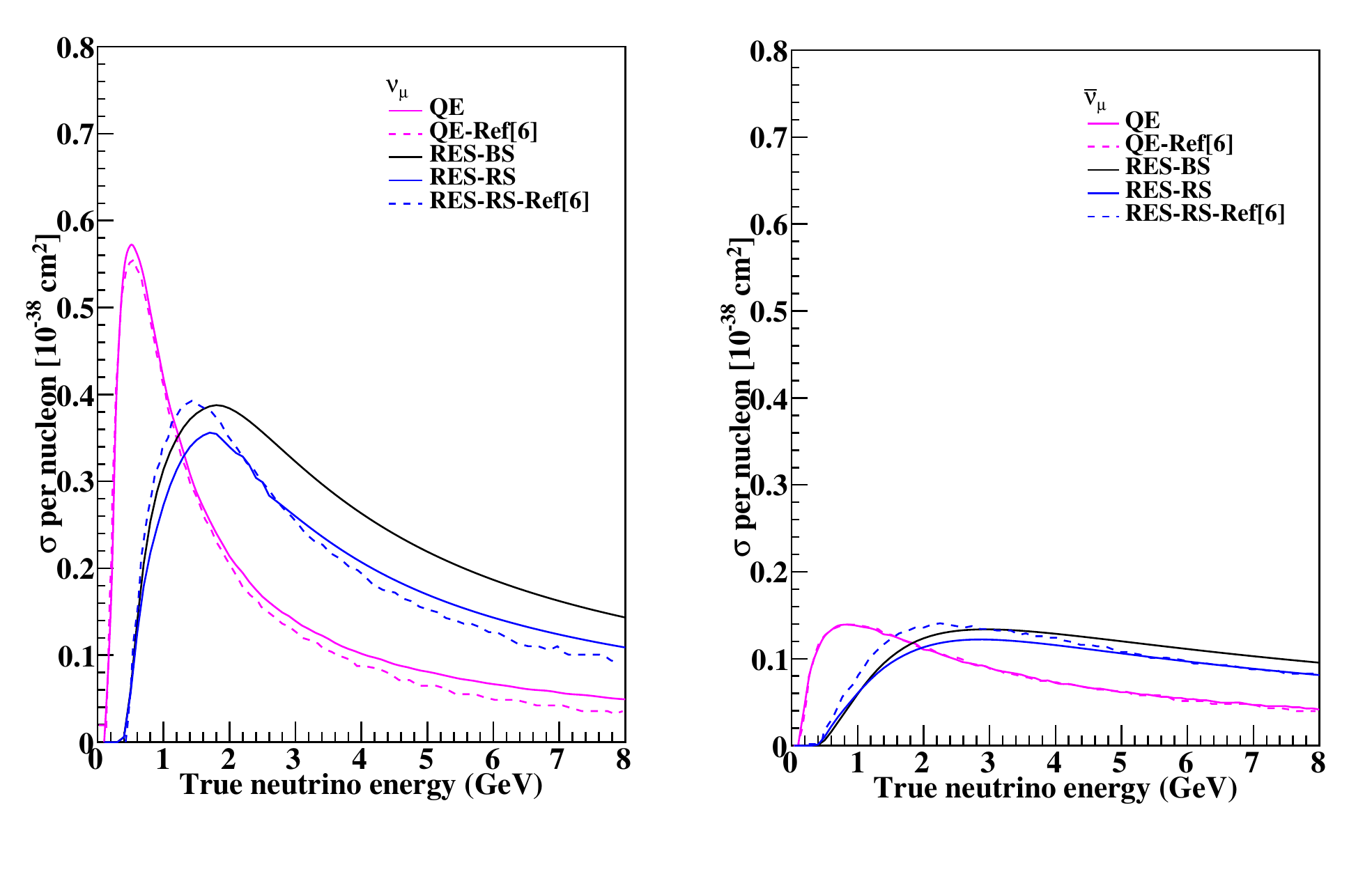}
\caption{Total CC cross-sections per nucleon divided by neutrino energy on carbon target  as a function of true neutrino energy for QE and RES processes for both neutrino (solid line) and antineutrino (dashed line). Comparisons are also shown for two models of RES processes: Berger-Sehgal and Rein-Sehgal. Here RES represents the $\Delta$-resonant pion production including the heavier resonance modes. The colour scheme for different cases: QE (magenta line), C12-BS (black line), and, C12-RS (blue line). Our cross-section results for QE, and RES-RS model for both $\nu_{\mu}$ and $\bar\nu_{\mu}$ are compared with those of  \cite{Ankowski:2015jya} (shown with dashed line, magenta for QE and  blue for RS). The labels QE and RES refer to quasi-elastic scattering and resonant pion production while BS and RS refer to Berger-Sehgal and Rein-Sehgal model.}
\end{figure}

In this paper, we explore the impact of nuclear effects (FSI) in resonance channels on the extraction of neutrino oscillation parameters at NO$\nu$A, for two nuclear models for RES processes, i.e. Rein-Sehgal and Berger-Sehgal models. We do this for the NO$\nu$A experiment (carbon target) for both neutrino and antineutrino for both the mass hierarchies using the GENIE neutrino event generator. Realistic detector specifications of NO$\nu$A, for different efficiencies, to see the variation in the sensitivity analysis is also included. We also compute the contribution of the pure QE interaction process to cross-section and event distribution and compare it with the contribution of resonance interactions, to show their relative importance. The paper is organised as follows. In Section II we briefly review some details of the NO$\nu$A experiment, for the sake of completeness of this work. Section III includes the principle of neutrino energy reconstruction, simulation, and physics details of this work and $\chi^{2}$ analysis followed by results and discussion in Section IV. We summarise our findings in Section V.

\section{The NO$\nu$A Experiment}
\label{sec:1}
The NO$\nu$A (NuMI Off-axis $\nu_{e}$ Appearance) \cite{NOvA:2016vij, NOvA:2017ohq} is a long base-line neutrino oscillation experiment designed to measure $\nu_{e}$ appearance and $\nu_{\mu}$ disappearance probability which uses high-intensity neutrino beam coming from Fermilab's Neutrinos at the Main Injector (NuMI) \cite{Adamson:2015dkw}. It consists of two functionally equivalent and identical detectors but different in dimension- 300 ton ND and 14 kton FD, both are placed at a long distance to measure neutrino oscillation. The ND is placed at a distance of 1 km downstream from the neutrino production source and it measures the un-oscillated neutrino beam as it is close to the source. In the ND the neutrinos do not get a chance to oscillate and it is ideal for cross-section measurement. As the ND is placed underground so the contribution of cosmic ray flux is negligible. The FD is situated at a distance of 810 km near Ash River, Minnesota to measure the oscillated neutrino flux. Both the detectors are placed 14.6 milli-radians off-axis from the center of the source beam as the NuMI beam generates a narrow band of neutrino energy spectra centered around 1.9 GeV in the FD at this position. One major advantage of the off-axis angle is that it suppresses the neutral current events and increases the number of CC $\nu_{\mu}$ signal events at this energy. The detectors are composed of cells of liquid scintillator encased in polyvinyl chloride (PVC) extrusions \cite{NOvA:2016kwd}. 

\begin{figure}[H]
 \centering\includegraphics[width=13.5cm, height=7cm]{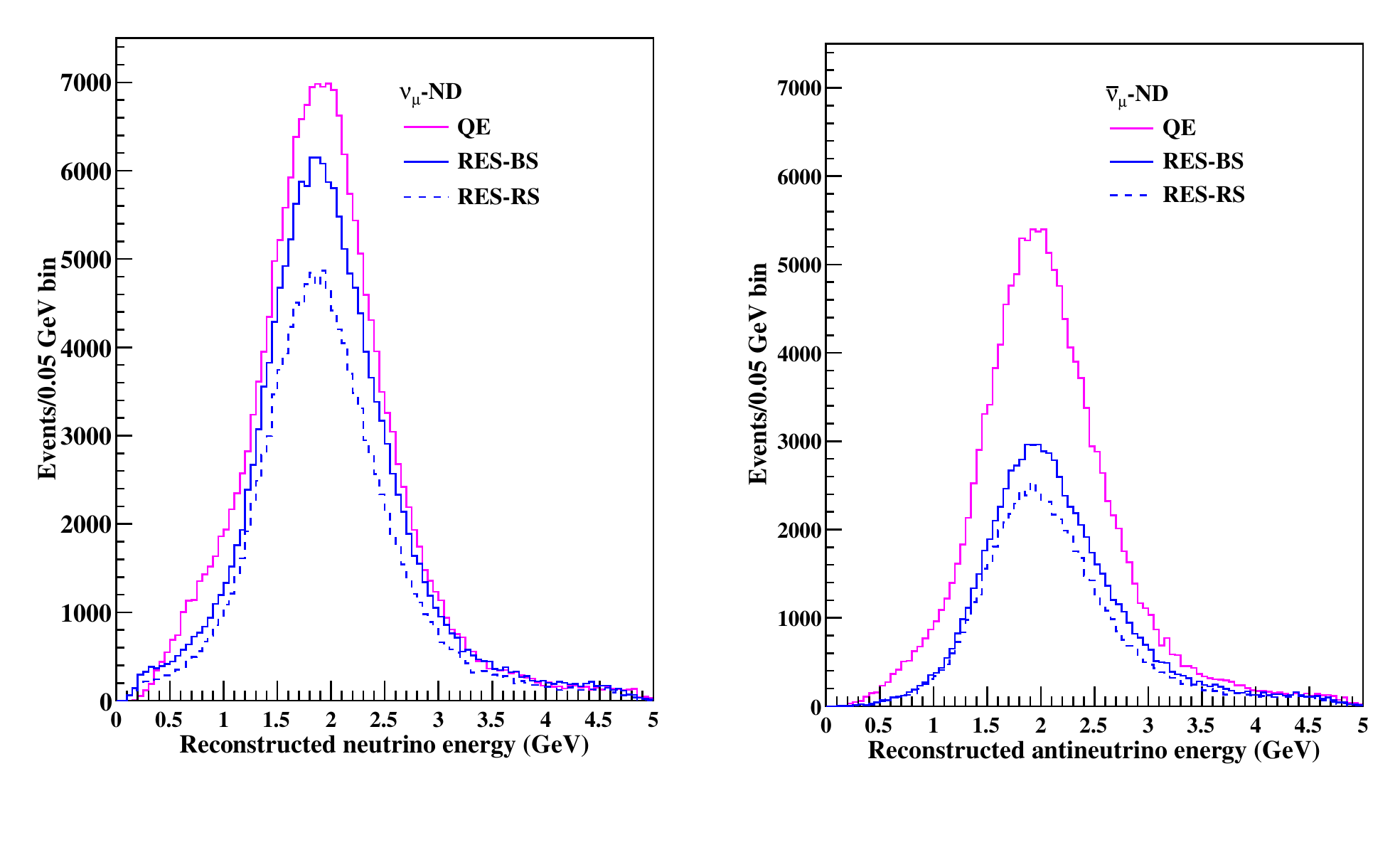}
 \caption{Comparison of ND events on carbon target as a function of calorimetric reconstruction energy for QE and RES processes for neutrino (right panel) and antineutrino (left panel). The colour scheme for different cases is - QE (solid magenta line), C12-BS (solid blue line), and, C12-RS (dashed blue line).}
\end{figure} 

\section{Physics and Simulation Details}
\label{sec:4}

The LBL neutrino experiments use event generators to generate events at the detectors for the simulation-based results. They play vital role in the design and execution of these experiments. GENIE (Generates Events for Neutrino Interaction Experiments) \cite{Andreopoulos:2009rq} is currently used by many major ongoing baseline experiments such as MINER$\nu$A, MINOS, MicroBooNE, NO$\nu$A, T2K, and DUNE. As NO$\nu$A currently employs the GENIE 2.12.2 t to simulate neutrino interactions in its ND and FD therefore among all other event generators we use a recent version of GENIE, v3.0.6 to investigate nuclear effects by generating cross-section, events, migration matrices and sensitivity contours. It simulates all types of relevant neutrino-nucleus interactions across the full kinematic space accessible at different types of neutrino oscillation experiments. Within GENIE, the neutrino-nucleus interaction simulation is separated distinctly - the initial nuclear state, the hard scatter, and the final state interactions of the produced particles within the nuclear medium. In the default GENIE configuration, the initial nuclear state is described by the global relativistic Fermi gas (RFG) model based on the work of Smith and Moniz \cite{Smith:1972xh} and modified by Bodek and Ritchie \cite{Bodek:1981wr} to account for the short-range nucleon-nucleon correlations by adding a high-momentum tail. The hard scattering is divided into four primary interaction types - QE, RES, DIS, and coherent (COH) interactions. COH interaction is a rare process where a neutrino scatters from the entire nucleus as a coherent whole. GENIE also simulates meson exchange currents (MEC), a process where the neutrino interacts with a nucleon coupled to another nucleon via a meson. For simulation, we have considered the QE, RES, two-nucleon knockout (2p2h), and deep inelastic scattering interaction processes. The default model for CC QE scattering is implemented using Llewellyn Smith \cite{LlewellynSmith:1971uhs} formalism. The pion production models are based on the phenomenological approach of Rein and Sehgal (RS model) \cite{Rein:1980wg} that describes pion production in the resonance region using nucleon-to-resonance transition matrix elements calculated using the relativistic quark model of Feynman-Kislinger-Ravndal (FKR model) \cite{Feynman:1971wr}. GENIE includes 16 resonances out of 18 resonances of the original RS paper in the region of transferred energy W < 1.7 GeV. RS model ignores the interference between neighboring resonances and does not include the lepton mass terms in the calculation of the differential cross-section, but it takes into account the impact of lepton mass on the phase space boundaries. The default value for the resonance axial vector mass $m_{A}$ is 1.12 GeV/$c^{2}$ in GENIE \cite{GENIE:2021npt}. Another updated model that is available for pion production is Berger and Sehgal (BS model) model \cite{Berger:2007rq} which is an extension of the RS model that includes the effects of lepton mass using the formalism of Kuzmin, Lyubushkin, and Naumov \cite{Kuzmin:2003ji}. The BS model reconsiders the FKR model applied to neutrino excitation of baryon resonances and also takes into account the pion-pole contribution to the hadronic axial vector current. For modelling of 2p2h processes, the Valencia model \cite{Nieves:2012yz} is used. The non-resonant background (NCB) is simulated by DIS contribution \cite{Paschos:2001np} with structure functions proposed by Bodek and Yang \cite{Bodek:2002ps}. We study both the RS and BS resonance models for our analysis and compare the cross-section, events, and sensitivity contours for the carbon target. The FSIs are modelled with INTRANUKE effective cascade model \cite{Ankowski:2009zz, Albrow:2007nu}. \\

\begin{figure}[H] 
\centering\includegraphics[width=13.5cm, height=7cm]{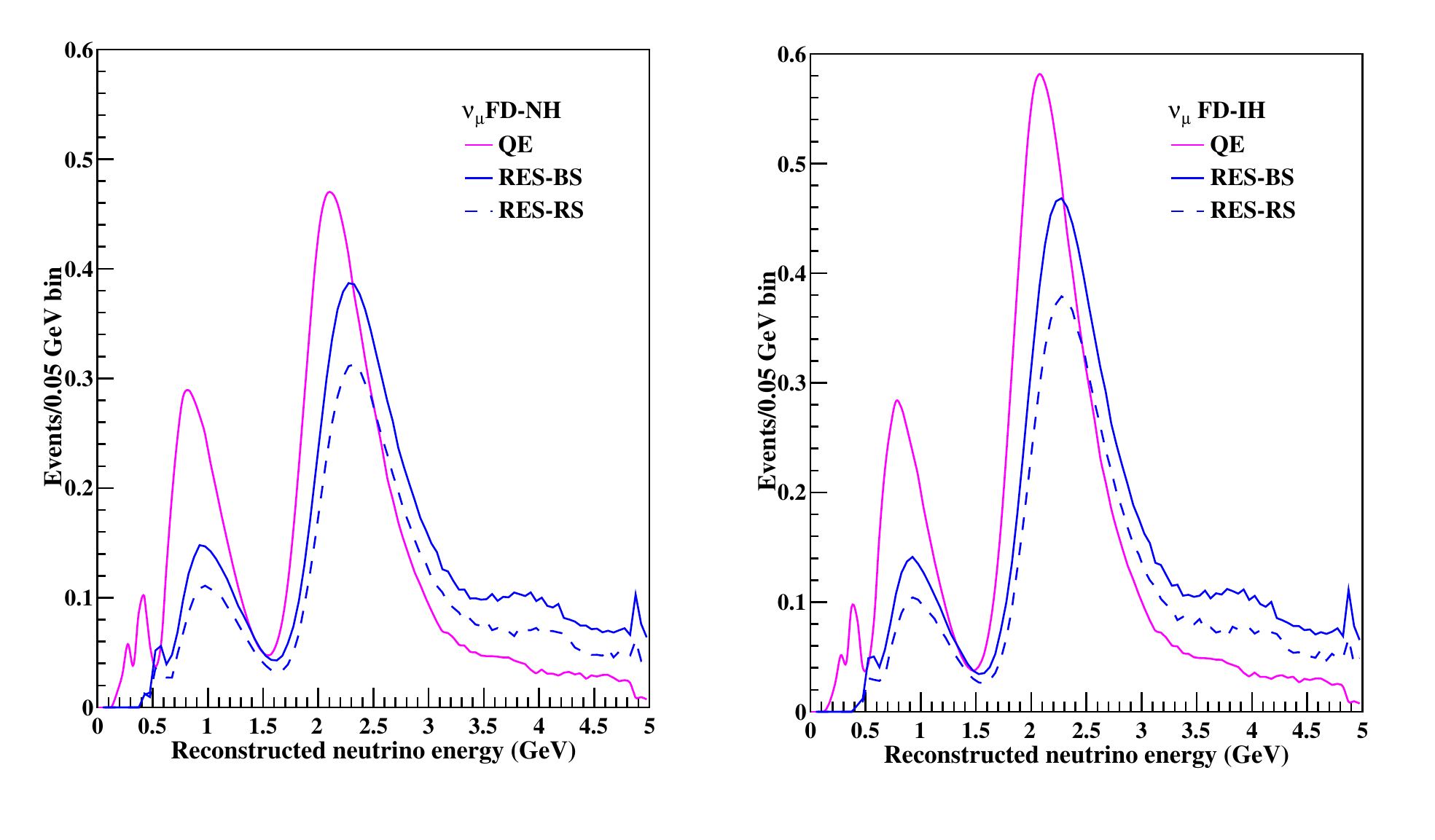}
\caption{Comparison of extrapolated FD events on carbon target as a function of calorimetric reconstruction energy for QE and RES processes for NH (right panel) and IH (left panel) for neutrino using two models of RES processes - Berger-Sehgal and Rein-Sehgal. The colour scheme is the same as Fig. 2.} 
\end{figure}

\begin{figure}[H] 
\centering\includegraphics[width=13.5cm, height=7cm]{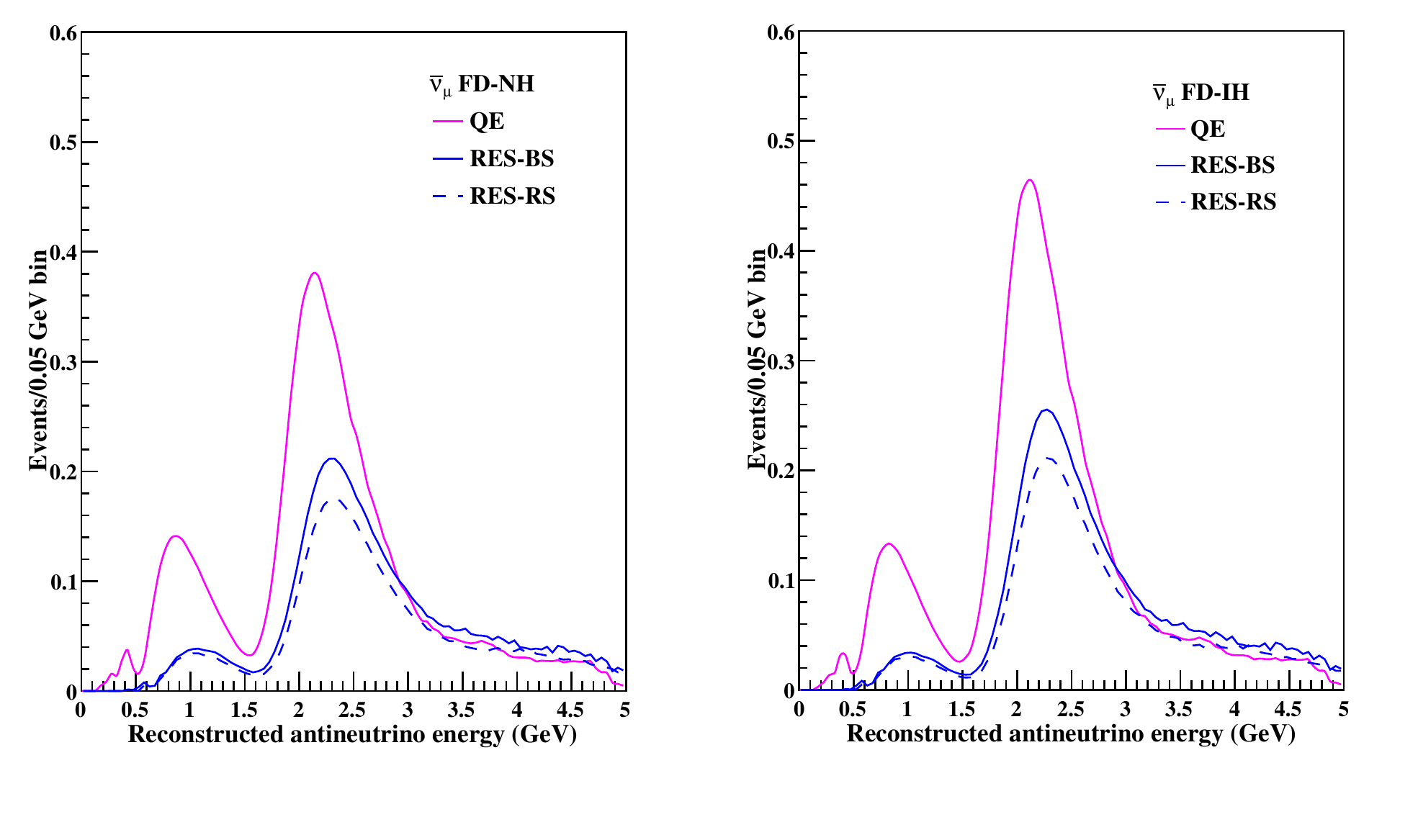}
\caption{Comparison of extrapolated FD events on carbon target as a function of calorimetric reconstruction energy for QE and RES processes for NH (right panel) and IH (left panel) for antineutrino. The colour scheme is the same as Fig. 2.} 
\end{figure}

For simulation, we generate an MC sample of half a million events using NO$\nu$A ND flux in the energy range 0-5 GeV for carbon target for both neutrino and antineutrino. The migration matrix for RES events is obtained using GENIE. For this work, only the $\nu_{\mu}\rightarrow\nu_{\mu}$ disappearance channel is considered. Neutral current (NC) interactions act as a background for this channel, however, we do not consider any background contribution in our analysis. We use equation (1) of \cite{Deka:2021qnw} to calculate the relevant oscillation probability for NO$\nu$A for the disappearance channel. The global best fit values of the oscillation parameters are taken from \cite{deSalas:2020pgw}. For neutrino energy reconstruction we use the calorimetric method, as recently we showed \cite{Deka:2022lvy} that this method produces more precision in sensitivity contours of neutrino oscillation parameters and this method relies on the capability of the detector design and reconstruction of final states. In this method, the neutrino energy is reconstructed by summing all the energies of the observed particles in the final state, and the direction of the outgoing particles are not needed. It is also defined as the summation of muon energy and hadronic energy. Muon energy is reconstructed from the length of the muon track in the detector and hadronic energy is measured by adding all the visible energy except the muon. The calorimetric reconstruction method can be applied to any type of CC interaction. In this method, the energy of a CC neutrino scattering off a nuclear target producing 'm' mesons and ejecting 'n' nucleons can be reconstructed as:

\begin{equation}
E^{cal}_{\nu}=E_{lepton}+\sum_{i}(E_{p}-M)+\epsilon_{n}+\sum_{m}E_{meson}
\end{equation}

where $E_{lepton}$ is the energy of the outgoing lepton, $E_{p}$ and $M$ define the energy of the $i$-th knocked-out nucleon and mass of the nucleon (target nucleus). Here, $\epsilon_{n}$ represents the single-nucleon separation energies of the outgoing nucleons and we keep it to a fixed value of 25 MeV for the carbon target for both neutrino and antineutrino. The same value of $\epsilon$ is added for every detected nucleon. $E_{meson}$ denotes the energy of the $m$-th produced meson. The expected number of FD events is determined using the extrapolation technique, in which the neutrino ND energy spectrum is extrapolated to the FD energy spectrum using a Monte Carlo simulation. The FD signal spectra of $\nu_{\mu}$($\bar\nu_{\mu}$) are predicted separately for neutrino and antineutrino beams based on the constrained $\nu_{\mu}$($\bar\nu_{\mu}$) event predictions in the ND. The advantage of this technique is that it can detect the smearing effect of imperfect energy resolution in ND and FD by creating a 2D histogram between the reconstructed energy and true energy in both the detectors. The full procedure of extrapolation can be found in Refs. \cite{Deka:2021qnw, NOvA:2016vij, NOvA:2017ohq, Deka:2022lvy, Pershey:2018gtf}. We use the Feldman-Cousins method \cite{Feldman:1997qc} to calculate the allowed confidence level in the parameter space, and the confidence regions are found from the condition:

\begin{equation}
\Delta \chi^{2}(\theta_{23}, \Delta m^{2}_{32}) \equiv \chi^{2}(\theta_{23}, \Delta m^{2}_{32})-\chi^{2}_{best-fit} < x
\end{equation}
where $x$= 2.30, 6.18, and 11.83 for the 1 $\sigma$, 2$\sigma$ and 3$\sigma$ confidence level, respectively. We also include detector response and efficiency in the analysis, where the measured energies are smeared with respect to the true values by finite detector resolution. Incorrect detector capabilities produce a non-vanishing probability such that an event with a true energy $E_{true}$ ends up being reconstructed with different energy $E_{rec}$. These non-vanishing probabilities are encoded in a set of migration matrices. We have considered realistic specification for NO$\nu$A - 31.2\% (33.9\%) efficiency of selection of $\nu_{\mu}$ ($\bar\nu_{\mu}$) events. For demonstration and comparison, we also analyzed 80\% detector efficiencies. 3.5\% muon energy resolution and 25\% hadron energy resolution respectively \cite{NOvA:2017ohq} is used, which gives an overall 7\% energy resolution for $\nu_{\mu}$-CC events for both detectors. 

\begin{table}[h]
\begin{center}
\begin{tabular}{|c|c|c|}
\hline
parameter & best fit & 3$\sigma$ range \\
\hline
$ \Delta m_{21}^2[10^{-5} eV^{2}]$ & $7.50$ & $6.94-8.14$\\
$ |\Delta m_{31}^2|[10^{-3} eV^{2}]$(NH) & $2.56$ & $2.46-2.65$ \\
$ |\Delta m_{31}^2|[10^{-3} eV^{2}]$(IH) & $2.46$ & $2.37-2.55$ \\
$ \sin^2\theta_{23}/10^{-1}$(NH) & $ 5.66$ & $4.46-6.09$ \\
$ \sin^2\theta_{23}/10^{-1}$(IH) & $ 5.66$ & $4.41-6.09$ \\
$ \sin^2\theta_{13}/10^{-2}$(NH) & $ 2.225$ & $2.015-2.417$ \\
$ \sin^2\theta_{13}/10^{-2}$(IH) & $ 2.250$ & $2.039-2.441$ \\
\hline
\end{tabular}
\end{center}
\caption{3$\sigma$ values of neutrino oscillation parameters taken from \cite{deSalas:2020pgw}, which are used in this work.}
\label{tab:data1}
\end{table}

\begin{table}[h]
\begin{center}
\begin{tabular}{|c|c|c|c|c|c|c|}
\hline
 & $\sin^{2}\theta_{23,min}$ & $\Delta m^{2}_{32,min}[eV^{2}]$ & $\chi^{2}_{min}$ & $\sin^{2}\theta_{23}$, $\Delta m^{2}_{32}$\scriptsize{(Global-fit)}  & Fig. no.\\
\hline
\scriptsize{$\nu_{\mu}$ RS (no detector) (NH)} & \scriptsize{$0.524$}  & \scriptsize{$2.56\times10^{-3}$} & \scriptsize{$0.937994$} & & \scriptsize{$(5,6)$}\\

\scriptsize{$\nu_{\mu}$ BS (no detector) (NH)} & \scriptsize{$0.524$}  & \scriptsize{$2.56\times10^{-3}$} & \scriptsize{$1.25265$}    &  & \scriptsize{$(5,6)$}\\

\scriptsize{$\nu_{\mu}$ RS (Eff=31.2\%) (NH)} & \scriptsize{$0.428$} & \scriptsize{$2.64\times10^{-3}$} & \scriptsize{$0.208237$} 	& \scriptsize{$0.566, 2.47\times10^{-3}$ (NH)} & \scriptsize{$(5)$}\\

\scriptsize{$\nu_{\mu}$ BS (Eff=31.2\%) (NH)} & \scriptsize{$0.476$} & \scriptsize{$2.56\times10^{-3}$} & \scriptsize{$0.332178$} 	&  & \scriptsize{$(5)$}\\

\scriptsize{$\nu_{\mu}$ RS (Eff=80\%) (NH)} & \scriptsize{$0.46$} & \scriptsize{$2.56\times10^{-3}$} & \scriptsize{$0.617576$} 	&  & \scriptsize{$(6)$}\\

\scriptsize{$\nu_{\mu}$ BS (Eff=80\%) (NH)} & \scriptsize{$0.572$} & \scriptsize{$2.64\times10^{-3}$} & \scriptsize{$0.883188$} 	&   & \scriptsize{$(6)$}\\

\hline
\scriptsize{$\nu_{\mu}$ RS (no detector) (IH)}  & \scriptsize{$0.54$}   & \scriptsize{$2.48\times10^{-3}$} & \scriptsize{$0.930316$}  &  & \scriptsize{$(5,6)$}\\

\scriptsize{$\nu_{\mu}$ BS (no detector) (IH)}  & \scriptsize{$0.54$}   & \scriptsize{$2.48\times10^{-3}$} & \scriptsize{$1.24795$}    &  & \scriptsize{$(5,6)$}\\

\scriptsize{$\nu_{\mu}$ RS (Eff=31.2\%) (IH)}  & \scriptsize{$0.572$} & \scriptsize{$2.56\times10^{-3}$}	 & \scriptsize{$0.208311$}   & \scriptsize{$0.566, 2.38\times10^{-3}$ (IH)}  & \scriptsize{$(5)$}\\

\scriptsize{$\nu_{\mu}$ BS (Eff=31.2\%) (IH)}  & \scriptsize{$0.572$} & \scriptsize{$2.56\times10^{-3}$}	 & \scriptsize{$0.274743$}   &  & \scriptsize{$(5)$}\\

\scriptsize{$\nu_{\mu}$ RS (Eff=80\%) (IH)}  & 	\scriptsize{$0.524$} & \scriptsize{$2.48\times10^{-3}$}	 & \scriptsize{$0.660938$}   &  & \scriptsize{$(6)$}\\

\scriptsize{$\nu_{\mu}$ BS (Eff=80\%) (IH)}  & 	\scriptsize{$0.572$} & \scriptsize{$2.56\times10^{-3}$}	 & \scriptsize{$0.833375$}   &  & \scriptsize{$(6)$}\\
\hline
\end{tabular}
\end{center}
\caption{Summary of best fit values and corresponding $\chi^{2}$ values of the oscillation parameters along with global best fit values for neutrino for different scenarios studied in this work.}
\label{tab:data1}
\end{table}

\begin{table}[h]
\begin{center}
\begin{tabular}{|c|c|c|c|c|c|c|}
\hline
 & $\sin^{2}\theta_{23,min}$ & $\Delta m^{2}_{32,min}[eV^{2}]$ & $\chi^{2}_{min}$ & $\sin^{2}\theta_{23}, \Delta m^{2}_{32}$\scriptsize{(Global-fit)}  & Fig. no.\\

\hline
\scriptsize{$\bar\nu_{\mu}$ RS (no detector) (NH)} & \scriptsize{$0.46$}  & \scriptsize{$2.56\times10^{-3}$} & \scriptsize{$0.435606$} &  & \scriptsize{$(7,8)$}\\

\scriptsize{$\bar\nu_{\mu}$ BS (no detector) (NH)} & \scriptsize{$0.46$}  & \scriptsize{$2.56\times10^{-3}$} & \scriptsize{$0.559778$}    & & \scriptsize{$(7,8)$}\\

\scriptsize{$\bar\nu_{\mu}$ RS (Eff=33.9\%) (NH)} & \scriptsize{$0.54$} & \scriptsize{$2.56\times10^{-3}$} & \scriptsize{$0.139485$} 	& \scriptsize{$0.566, 2.47\times10^{-3}$ (NH)} & \scriptsize{$(7)$}\\

\scriptsize{$\bar\nu_{\mu}$ BS (Eff=33.9\%) (NH)} & \scriptsize{$0.572$} & \scriptsize{$2.64\times10^{-3}$} & \scriptsize{$0.184904$} 	& & \scriptsize{$(7)$}\\

\scriptsize{$\bar\nu_{\mu}$ RS (Eff=80\%) (NH)} & \scriptsize{$0.444$} & \scriptsize{$2.64\times10^{-3}$} & \scriptsize{$0.281882$} 	& & \scriptsize{$(8)$}\\

\scriptsize{$\bar\nu_{\mu}$ BS (Eff=80\%) (NH)} & \scriptsize{$0.428$} & \scriptsize{$2.64\times10^{-3}$} & \scriptsize{$0.41076$} 	&   & \scriptsize{$(8)$}\\

\hline
\scriptsize{$\bar\nu_{\mu}$ RS (no detector) (IH)}  & \scriptsize{$0.54$}   & \scriptsize{$2.48\times10^{-3}$} & \scriptsize{$0.930316$}  &  & \scriptsize{$(7,8)$}\\

\scriptsize{$\bar\nu_{\mu}$ BS (no detector) (IH)}  & \scriptsize{$0.54$}   & \scriptsize{$2.48\times10^{-3}$} & \scriptsize{$0.571568$}    &  & \scriptsize{$(7,8)$}\\

\scriptsize{$\bar\nu_{\mu}$ RS (Eff=33.9\%) (IH)}  & 	\scriptsize{$0.428$} & \scriptsize{$2.56\times10^{-3}$}	 & \scriptsize{$0.10904$} &  \scriptsize{$0.566, 2.38\times10^{-3}$(IH)}  & \scriptsize{$(7)$}\\

\scriptsize{$\bar\nu_{\mu}$ BS (Eff=33.9\%) (IH)}  & \scriptsize{$0.588$} & \scriptsize{$2.56\times10^{-3}$}	 & \scriptsize{$0.173555$}   & & \scriptsize{$(7)$}\\

\scriptsize{$\bar\nu_{\mu}$ RS (Eff=80\%) (IH)}  & 	\scriptsize{$0.572$} & \scriptsize{$2.56\times10^{-3}$}	 & \scriptsize{$0.240326$}   & & \scriptsize{$(8)$}\\

\scriptsize{$\bar\nu_{\mu}$ BS (Eff=80\%) (IH)}  & 	\scriptsize{$0.428$} & \scriptsize{$2.56\times10^{-3}$}	 & \scriptsize{$0.341294$}  &  & \scriptsize{$(8)$}\\
\hline
\end{tabular}
\end{center}
\caption{Summary of best fit values and corresponding $\chi^{2}$ values of the oscillation parameters along with global best fit values for antineutrino for different scenarios studied in this work.}
\label{tab:data1}
\end{table}

\section{Results and Discussion}
\label{sec:5}

In this section, we present the results of this study, i.e., the impact of the nuclear effect (RES processes) on cross-section, events, and oscillation analysis for two nuclear models. In Fig. 1, we show the comparison of the contribution of $\Delta$-resonant pion production along with heavier resonance modes to the cross-section of carbon target for two different nuclear models of RES. We have also shown the contribution of the QE process to the cross-section, to show the relative importance of the two interaction processes. For NO$\nu$A neutrino flux below 1 GeV, QE scattering is the main channel of neutrino-nucleus interaction but as the neutrino energy increases (> 2GeV), RES contribution to the cross-section increases significantly and it becomes the dominant interaction process. This may lead to the production of many hadrons in the final state which will impact the reconstruction of neutrino energy. Also for BS and RS resonance models, there is a non-negligible difference between them which reflects the lack of understanding of theoretical models. These differences are small in the case of antineutrinos as compared to the neutrino. However, these differences can not be ignored as they will propagate to the determination of neutrino oscillation parameters. Same observations are also seen in ND events, FD events, and sensitivity analysis. We have also shown the comparison of our cross-section results of QE and RES-RS with Ref. \cite{Ankowski:2015jya}, and we note that agreement between the two is very good. \\

In Fig. 2, the binned ND events as a function of the reconstructed neutrino energy are shown in the left panel for neutrino and right for antineutrino when carbon is used to generate the corresponding cross-sections. From Fig. 2, we observe RES contribution to event distribution is large in comparison with QE contribution in the case of neutrino as compared to antineutrino due to differences in the cross-sections. There is a clear distinction between the BS and RS models and this difference is less for antineutrino as also seen in Fig. 1. The difference between both the models is less in the lower and upper energy range but in the peak region at around 2 GeV the differences are visible for both neutrino and antineutrino. Similar observations can also be noticed in FD event distributions. The difference between the models further propagates to the FD event distribution which further transforms into a sizeable effect for the oscillation analysis. The number of expected extrapolated RES events binned as a function of calorimetric reconstruction of neutrino energy at the FD for different contributions is shown in Fig. 3 (for neutrino) and Fig.4 (for antineutrino) for both mass hierarchies-normal hierarchy (NH, left panel) and inverted hierarchy (IH, right panel). The $E^{cal}_{\nu}$ distributions of these events are obtained by using the carbon migration matrices. 

\begin{figure}[H] 
 \centering\includegraphics[width=13cm, height=10cm]{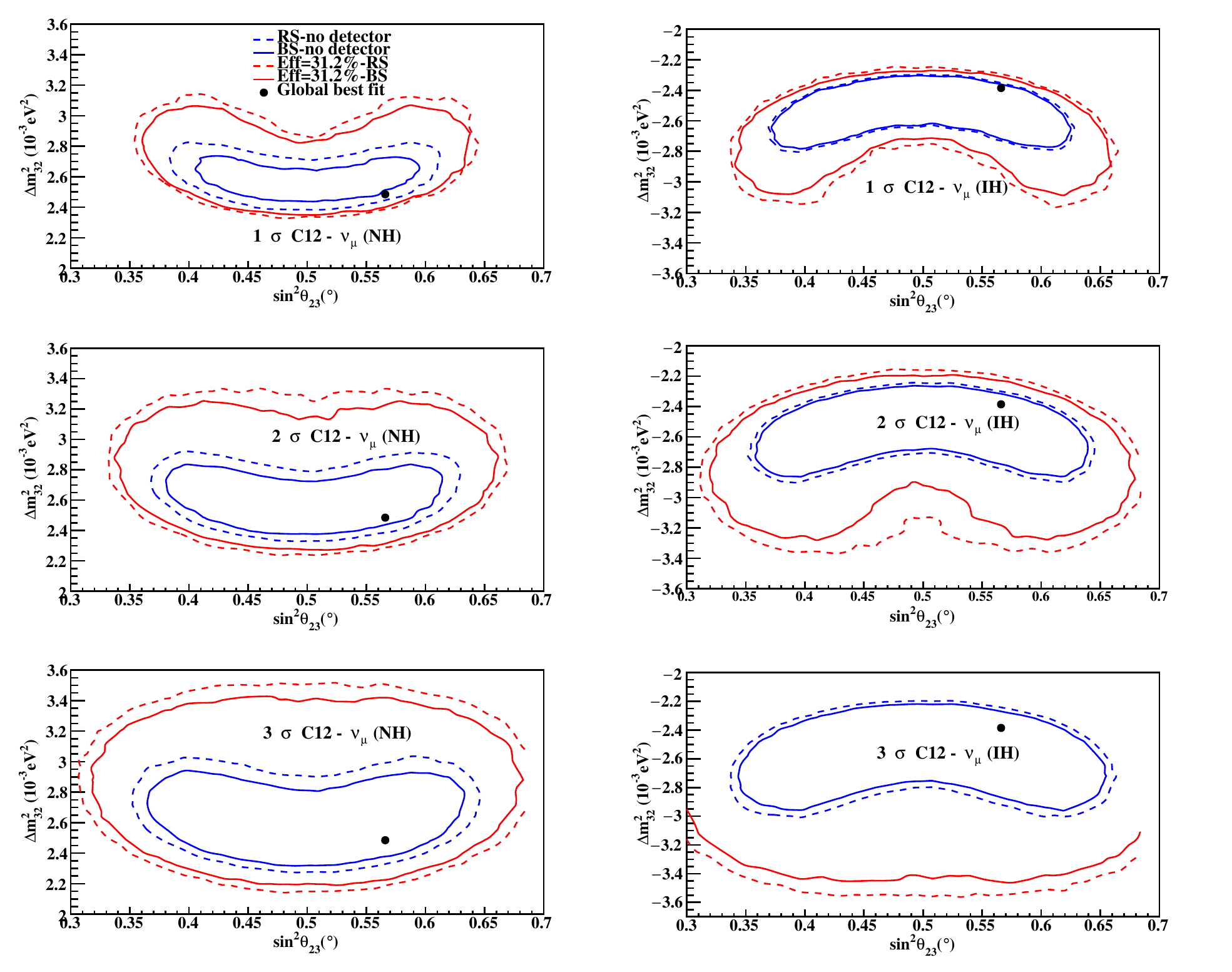}
\caption{Comparison of 1$\sigma$, 2$\sigma$ and 3$\sigma$ confidence regions in the $\theta_{23}$-$\Delta m^{2}_{32}$ plane using carbon target for RES processes for NH (right panel) and IH (left panel) for neutrino with/without detector effect. The color scheme for different cases: C12-BS (solid red line, 31.2\% efficiency), C12-RS (dashed red line, 31.2\% efficiency), C12-BS (solid blue line, no detector effect), and Ar40-RS (dashed blue line, no detector effect). The black dot represents the value of the $\chi^{2}$ at the global best fit.} 
\end{figure}

\begin{figure}[H] 
\centering\includegraphics[width=13cm, height=10cm]{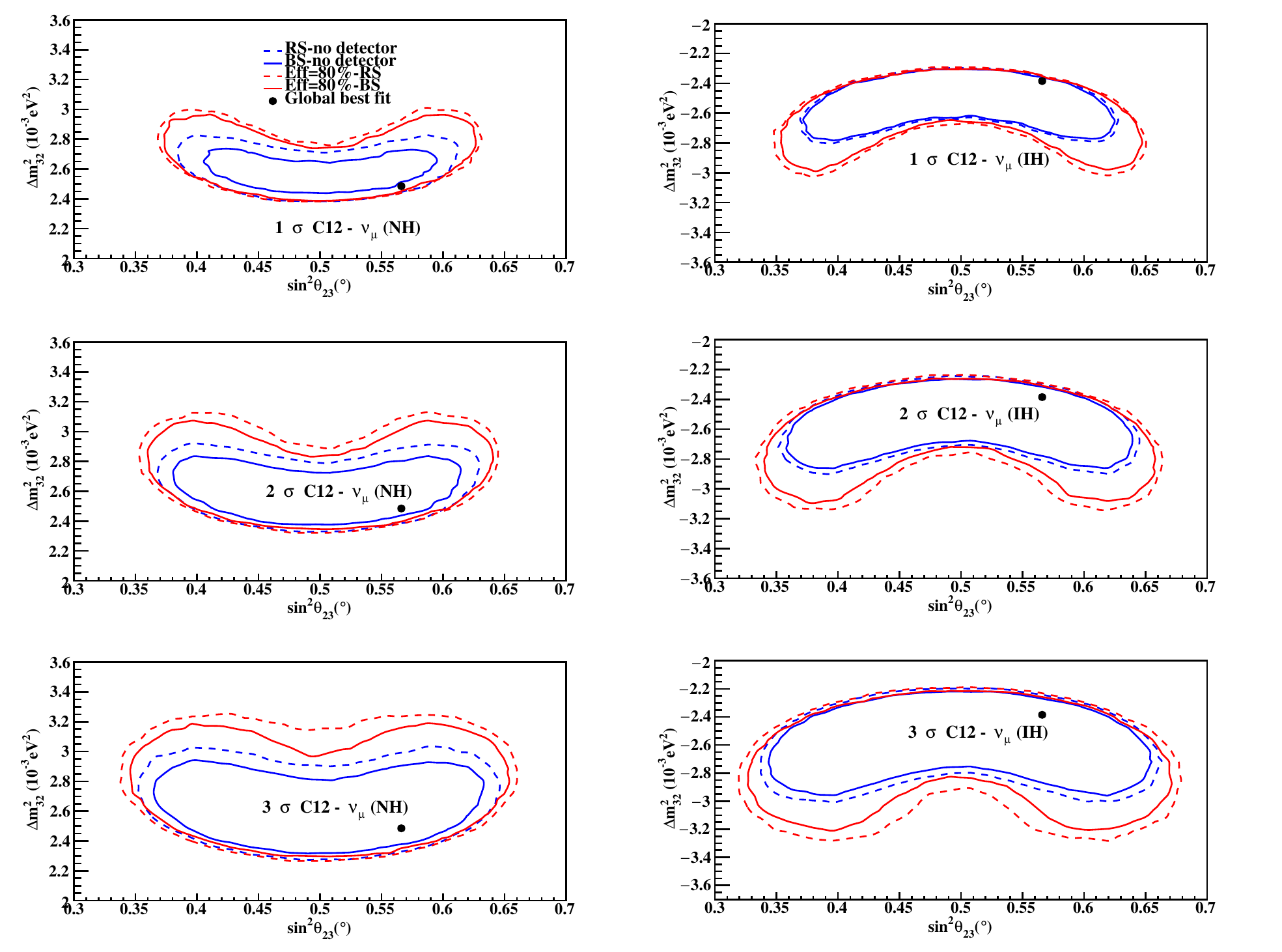}
\caption{Same as Fig.5 but for 80\% detector efficiency.} 
\end{figure}

\begin{figure}[H] 
\centering\includegraphics[width=13cm, height=10cm]{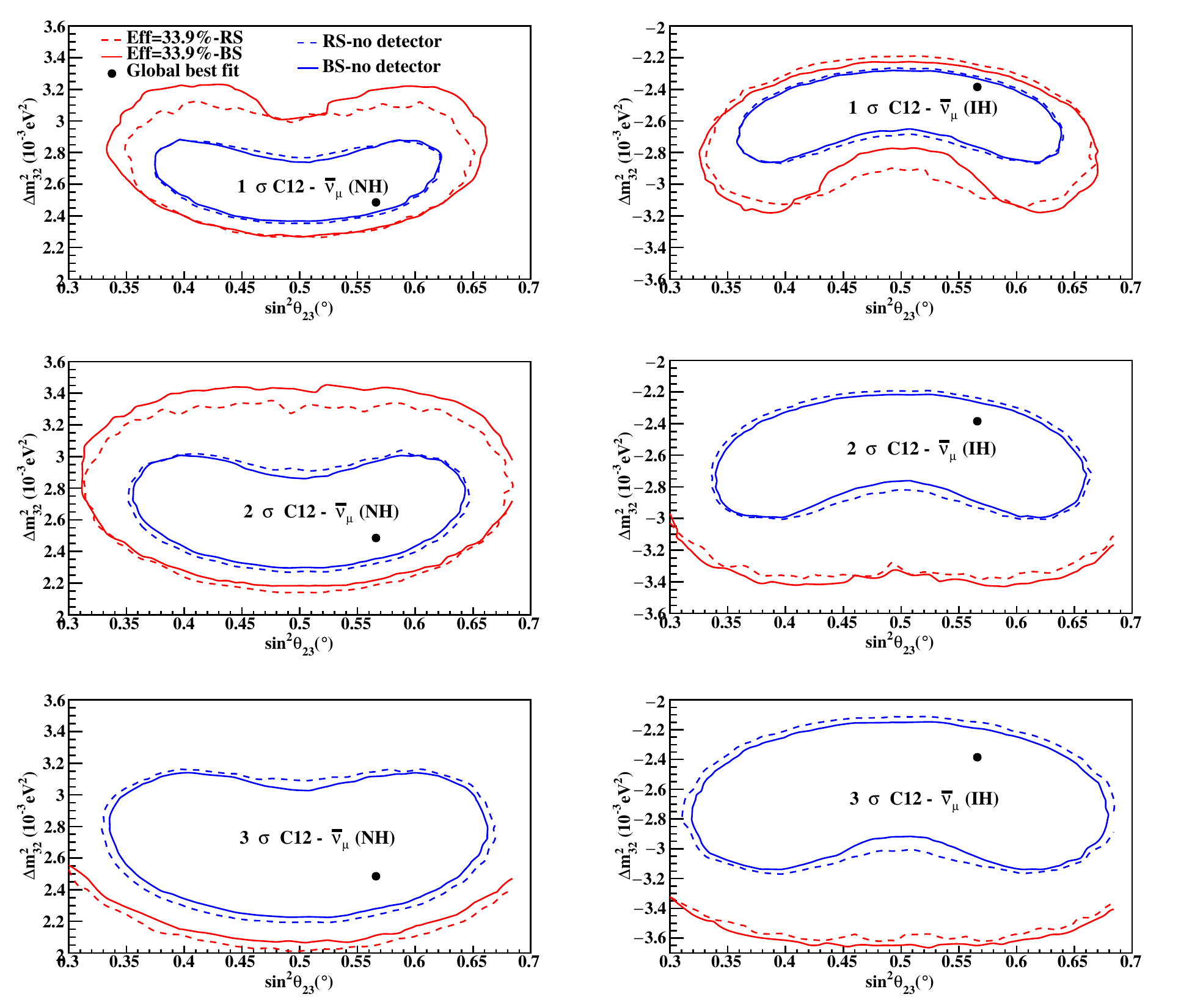}
\caption{Same as Fig. 5 but for antineutrino and detector efficiency 33.9\%.} 
\end{figure}

\begin{figure}[H] 
\centering\includegraphics[width=13cm, height=10cm]{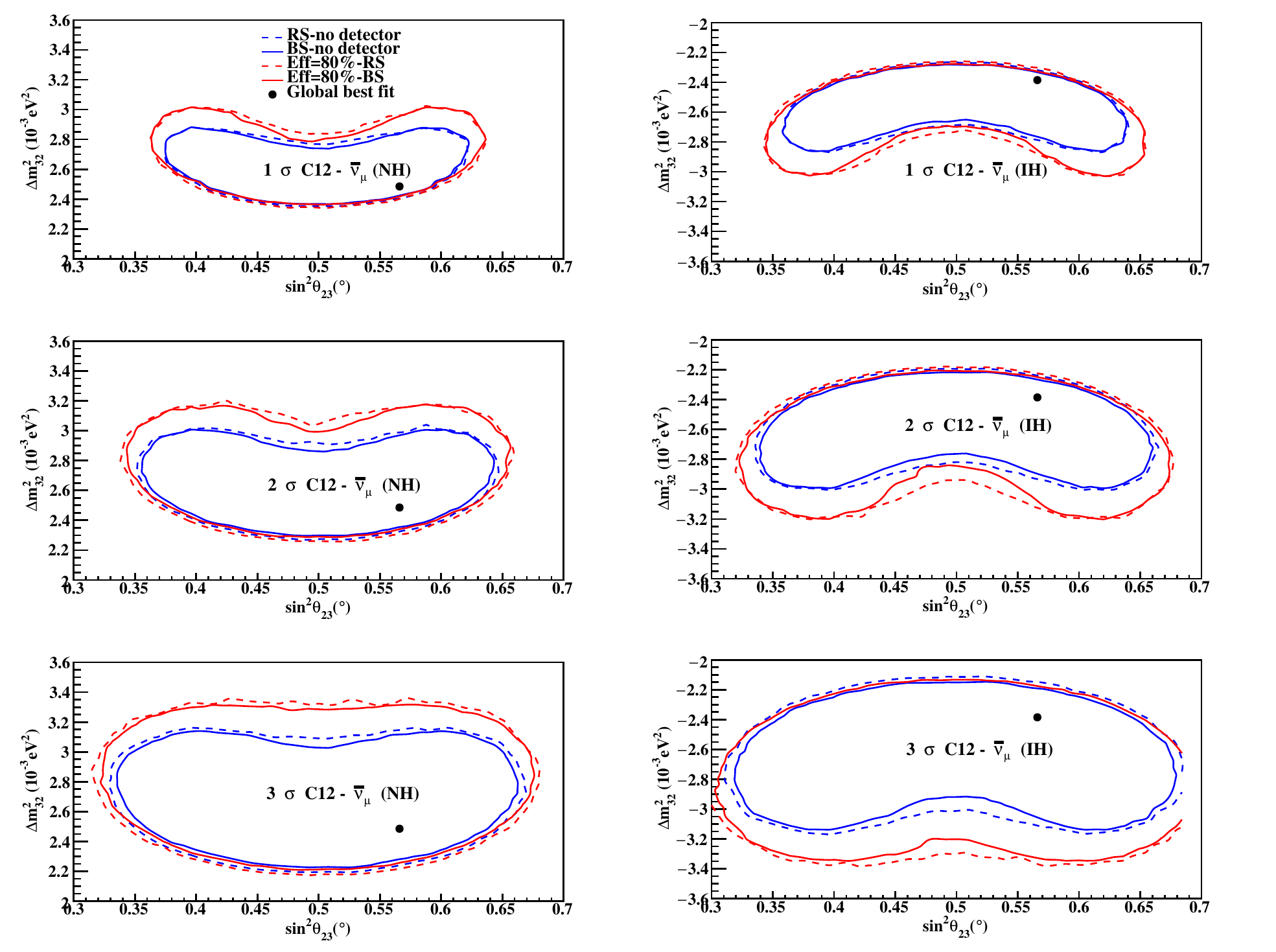}
\caption{Same as Fig. 5 but for antineutrino and detector efficiency 80\%.} 
\end{figure}

In Figs. 5-8, we represent the comparison of 1$\sigma$, 2$\sigma$ and 3$\sigma$ sensitivity contours in the $\theta_{23}$-$\Delta m^{2}_{32}$ plane for different cases: no detector-BS (solid blue line), no detector (dashed blue line), efficiency=31.2\%(33.9\% or 80\% )-BS (solid red line), and efficiency=31.2\%(33.9\% or 80\%)-BS (dashed red line) for both neutrino and antineutrino for both NH and IH. We show how the detector response changes when we overestimate the detector efficiency by 80\%. The back dot indicates the location of the best-fit point. These contour areas represent the analysis where nuclear effects are taken into account by using the carbon migration matrices to calculate the fitted rates. Comparing the area inside the confidence levels, we observe that out of the RS and BS models, the BS model shows a better precision in the measurement of neutrino oscillation parameters as it has a lesser area inside the contour than the RS. This observation is as expected as observed from the ND and FD events in which it is seen that the number of events is more in BS than RS. These differences due to nuclear effects can be corrected for oscillation analysis using a proper nuclear model and Figs. 5-8 also show the importance of developing an accurate theoretical description of nuclear effects for the targets used in neutrino oscillation experiments. In the case of antineutrino in Fig. 7 and 8, the difference between the contour areas for BS and RS is less as compared to neutrino as expected from Figs. 1-4, and also for IH, the difference between the two models is less as compared to NH. From Figs. 5-8, it is observed that confidence levels with 80\% detector efficiency are more close to no detector effect (which corresponds to 100\% efficiency and no resolution function for the detector) compared 31.2\% ($\nu_{\mu}$) or 33.9\% ($\bar\nu_{\mu}$) efficiency. This is also as expected, as better detector resolution will reduce uncertainty. We have also shown the $\chi^{2}$ values  of the oscillation parameters for each case with/without detector effects and shown the deviation from the global best fit values of $\sin^{2}\theta_{23}$ and $\Delta m^{2}_{32}$. Table II is for neutrino while Table III is for antineutrino.

\section{Summary and Conclusion}
\label{sec:5}

In this work, we extensively analysed the impact of resonance processes in neutrino-nucleus interaction on the sensitivity analysis of neutrino oscillation parameters- $\theta_{23}$ and $\Delta m^{2}_{32}$ in NO$\nu$A experiment, using two models - Berger-Sehgal and Rein-Sehgal for carbon target in the $\nu_{\mu}\rightarrow\nu_{\mu}$ disappearance channel. We observed a sizeable amount of uncertainties present in the models, for both neutrino and antineutrino for both the mass hierarchies. We also included the realistic detector setup specifications of NO$\nu$A, and analysed two detector efficiencies (31.2\% or 33.9\% and 80\%), for the sake of illustration and comparison. We compared  our cross-section results of QE and RES-RS with those of Ref. \cite{Ankowski:2015jya}, and the agreement between the two is found to be very good. It is observed  that RES processes have a significant contribution to cross-section, events, and sensitivity analysis. The Berger-Sehgal model gives more precision than the Rein-Sehgal model on the extraction of neutrino oscillation parameters. Also, sensitivity contours with 80\% detector efficiency were found to be more close to no detector analysis. The lower efficiency of the detector implies more uncertainties which indicate the need for improvement in detector efficiencies in future experiments. The NO$\nu$A flux peaks around 2 GeV and in this range resonance process is the dominant process as seen in Fig. 1, therefore it should be handled carefully to reduce uncertainties. As the resonance process contributes significantly to the neutrino-nucleus cross-section, so it is very important to understand the nuclear effects correctly for the construction of a nuclear model which will reduce the theoretical uncertainties in the nuclear models. Obtained results show that an accurate description of nuclear effects, i.e. RES processes studies here, contributes significantly to the determination of neutrino oscillation parameters in oscillation experiments. This study will help to quantify the systematic uncertainties present in the models, so that they may be included carefully in simulation tools to increase precision in neutrino parameter measurements in the future.

\section{Acknowledgments} 
\label{sec:6}

PD would like to thank Jaydip Singh for useful discussions.

\section{Declarations}
\textbf{Conflicts of interests} The authors declare no potential conflict of interests.

\end{document}